\documentclass[aps,prl,twocolumn,superscriptaddress,showpacs]{revtex4}
 \usepackage{amsmath}
\usepackage{graphicx}
\usepackage{epsfig}
\newcommand{\beq}{\begin{equation}}
\newcommand{\eeq}{\end{equation}}
\newcommand{\bea}{\begin{eqnarray}}
\newcommand{\eea}{\end{eqnarray}}

\begin{document}
\bibliographystyle{apsrev}
 
\title{Hall effect between parallel quantum wires }
\author{  M. Kindermann}
\affiliation{ School of Physics, Georgia Institute of Technology, Atlanta, Georgia 30332, USA   }

\date{ July 2007}
\begin{abstract}
 We study theoretically the parallel  quantum wires of the experiment by Auslaender {\it et al.} [Science {\bf 308}, 88 (2005)] at low electron density.     It is shown that a   Hall effect as observed in two- or three-dimensional electron systems develops  as one of the two wires enters the spin-incoherent regime of small spin bandwidth.  This together with magnetic field dependent tunneling exponents clearly identifies spin-incoherence in such experiments and it serves to distinguish it from disorder effects.  
 \end{abstract}
\maketitle
Since its discovery over a century ago the Hall effect has helped to uncover  a number of most fundamental physical effects.  Among the most famous  are the quantization of electrical conductance in the integer quantum Hall effect \cite{klitzing:prl80}, the fractionalization of electric charge in the fractional quantum Hall effect \cite{tsui:prl82}, and anomalous velocities due to Berry phases in   ferromagnets \cite{pugh:rmp53,niu:prb99}. 

In this Letter we show that, contrary to what one may expect, Hall measurements are also a powerful probe of one-dimensional quantum wires. We predict clear signatures of ``spin-incoherent'' physics in Hall measurements on tunnel-coupled, parallel quantum wires. 
The spin-incoherent limit of the interacting one-dimensional electron gas is reached when the temperature $T$ becomes larger than the spin bandwidth $J$, $kT \gg J$. This regime is a generic property of  interacting electrons at low densities, when a Wigner crystal with large inter-electron spacing is formed. As one of the few known regimes of one-dimensional conductors that displays physics qualitatively different from the conventional Luttinger liquid this limit has received much recent theoretical attention \cite{fiete:rmp07,kindermann07,fiete07}.   Experimentally, however, it has not been identified conclusively, yet.  One of the most promising candidate systems for reaching the low density regime required for observing spin-incoherent physics are the semi-conductor quantum wires of the experiment by  Auslaender {\it et al.}, Refs.\ \cite{auslaender:sci02,auslaender:sci05}.  The tunneling current in that experiment  has shown a loss of momentum resolution at low electron densities. This finding was likely due to a breaking of translational invariance by disorder \cite{auslaender:sci05}, but it is also the main previously known   \cite{fiete:prb05b} signature of spin-incoherence in the experimental arrangement of Refs.\ \cite{auslaender:sci02,auslaender:sci05}.    An experimental probe that is able to distinguish spin-incoherent physics from the breaking of translational invariance in that experimental setting is thus urgently needed if spin-incoherence is to be observed in such experiments. The Hall measurements proposed here are such a probe \cite{footnote00}.
\begin{figure} 
\includegraphics[width=6cm]{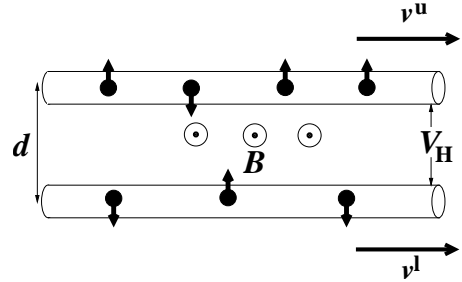}
\caption{    Two tunnel-coupled, one-dimensional wires at a distance $d$ in a perpendicular magnetic field $B$. At low densities the conduction electrons form Wigner crystals. The crystals are sliding at velocities $v^{\rm u}$ and $v^{\rm l}$ when  electrical currents flow. The figure illustrates wires at $J\ll kT$. They have an effectively static spin configuration and an almost conventional Hall voltage $V_{\rm H}$ appears.}   \label{fig1}
\end{figure} 

In the experiments of Refs.\ \cite{auslaender:sci02,auslaender:sci05} two parallel one-dimensional wires in a perpendicular magnetic field $B$ are close enough for electrons to tunnel between them, see Fig.\ \ref{fig1}.  A Hall effect in this geometry should induce a voltage $V_{\rm H}$ {\em between} the two wires in response to a current $I$ flowing {\em through} them. For noninteracting electrons in a  translationally invariant setup, however, no such voltage is expected. Tunneling then is momentum-resolved and occurs only between a few discrete momentum states. In the generic case that the current $I$ that flows {\em through} the wires is not carried by any of the states that participate in the tunneling {\em between} them,    the tunnel current, and correspondingly $V_{\rm H}$, is independent of $I$.  Nevertheless, a transverse voltage can be observed in such experiments if  translational invariance is broken or through electron-electron interactions. 
We show that at $kT\ll J$ the breaking of translational invariance induces a transverse voltage $V_{\rm H}$ that is generically  weak and very unconventional in that it is nonlinear in  $B$. In contrast, in the spin-incoherent regime of $kT \gg J$ a Hall effect as known from higher-dimensional electron systems is found, with a Hall voltage linear in $B$ and $I$.  This clear signature of spin-incoherence, distinguishing it from disorder effects, makes Hall measurements on parallel quantum wires a promising tool in the search for this new and exciting type of one-dimensional physics.

The emergence of traditional Hall physics in  spin-incoherent Wigner crystals is due to the nearly classical character of charge transport in this regime. When electrical currents $I^\mu$  flow the Wigner crystals  are sliding  at  velocities $v^\mu\propto I$.  Here, the index $\mu\in\{{\rm u,l}\}$ distinguishes the upper from the lower wire in Fig.\ \ref{fig1}.  At $kT \gg J^\mu$ the electrons on the lattice sites of the crystal are distinguishable through the effectively static spins  attached to them    and  therefore  behave very similarly to classical, charged particles. They   experience a Lorentz force $\propto I$ that 
induces   an (almost) conventional Hall voltage.

{\em Calculation:}  To lowest order in the tunnel coupling $\lambda$  between the  wires of a setup as shown in Fig.\ \ref{fig1} the  tunneling current $I_{\rm T}$ between them takes the form \cite{tserkovnyak:prb03}
\begin{widetext}
\bea \label{tunnel}
  I_{\rm T} &=& e |\lambda|^2 \sum_{\sigma }  \int dt dx dx'\, e^{ i eV_{\rm T}t+i q_{\rm B}(x-x')}
  \left[ G^>_{{\rm u}\sigma}(x,x',t)G^<_{{\rm l}\sigma}(x',x,-t)-
  G^<_{{\rm u}\sigma}(x,x',t)G^>_{{\rm l}\sigma}(x',x,-t)\right]. \label{eq:Iint}
\eea
\end{widetext}
Here, $V_{\rm T}  $ is the difference between the chemical potentials of the wires (we set $\hbar=1$). In a magnetic field $B$ the electrons experience a momentum boost $q_{\rm B}=eBd$ when tunneling  between the wires that are a distance $d$ from each other \cite{footnote0}. $G_{\rm u}$ and $G_{\rm l}$ are the electron Green functions in the upper and the lower wire respectively. They depend on the    currents $I^\mu$  that flow through the  wires.

\begin{figure} 
\includegraphics[width=8.5cm]{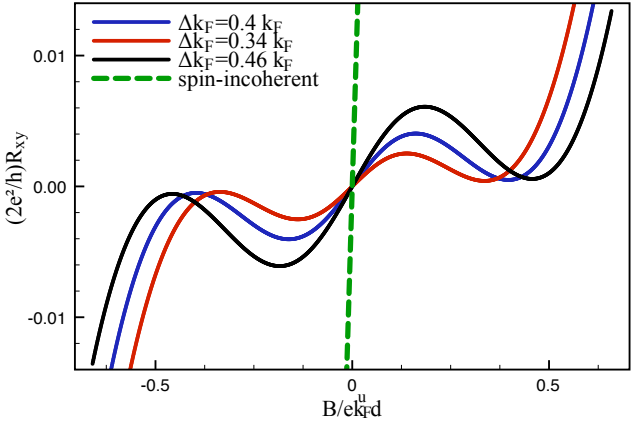}
\caption{ Transverse resistance $R_{\rm xy}$ of two coupled quantum wires at $I^{\rm u}=I^{\rm l}$.  At $kT \ll J^\mu$ (solid lines) the dependence on $B$   is nonlinear. In  the spin-incoherent case $kT\gg J^\mu$ (broken line), in contrast,  $R_{xy}$   is linear in $B$ with a slope  greatly exceeding $dR_{xy}/dB|_{B=0} $  at $kT \ll J^\mu$ (solid line: $\Delta k_{\rm F} l_{\rm br} \gg 1$; broken line:   for  identical wires).   }   \label{fig2}
\end{figure} 

{\em Broken translational invariance:} We first consider the case that translational invariance is broken, but $kT\ll J^\mu$, such that the wires have not entered the spin-incoherent regime.  At sufficiently low energies such wires are described by   Luttinger liquids \cite{haldane:jpc81} with Fermi wavevectors  $k_{\rm F}^{\mu}$, Fermi velocities $v_{\rm F}^\mu$  and interaction parameters $g^{\mu}_{\rm c}$ and $g^{\mu}_{\rm s}$ of their charge and  spin modes respectively  \cite{giamarchi:bo04}.   We assume that translational invariance is broken over a length $l_{\rm br}$ that is shorter than    the electron wavepackets     such that $l_{\rm br}$ shows in observables,  $eV, kT \ll v_{\rm F}^\mu/l_{\rm br}$, where $V={\rm max}\{V_{\rm T}, I^{\rm u}/e, I^{\rm l}/e\}$.  In the experiments of Refs.\ \cite{auslaender:sci02,auslaender:sci05} momentum conservation is typically lifted through the finite length of the tunneling region, disorder, or a leakage of electrons into the surrounding two-dimensional electron gas  with mean free path $l_{\rm 1D-2D}$. We first assume  that the latter is the dominant mechanism, such that    $l_{\rm br}=l_{\rm 1D-2D}$.  At   $eV \ll  kT, v_{\rm F}^\mu |q_{\rm B}\pm k_{\rm F}^{\rm u}\pm k_{\rm F}^{\rm l}|$   we then   find  
\bea \label{LL}
I_{\rm T}&\propto& T^\alpha\sum_{\sigma^{\rm u},\sigma^{\rm l}=\pm1} f\left(\sigma^{\rm u}k_{\rm F}^{\rm u}+\sigma^{\rm l}k_{\rm F}^{\rm l}-q_{\rm B}\right) \\
 &&\;\;\;\;\;\;\;\;\;\;\;\;\;\;\;\;\;\;\mbox{} \times\left( \pi \frac{ \sigma^{\rm u}I^{\rm u}+\sigma^{\rm l}I^{\rm l}}{2e^2}-V_{\rm T}\right)\nonumber
 \eea
with  $\alpha=-1+\sum_{\nu\in\{{\rm c,s}\}} (g^{\rm u}_\nu+g^{\rm u-1}_\nu+g^{\rm l}_\nu+g^{\rm l-1}_\nu)/4$ and $f(k) = l_{\rm br}/(1+k^2 l^2_{\rm br})$.  The transverse voltage $V_{\rm H}$ is found as the counter voltage $V_{\rm H}=-V_{\rm T}$  needed to cancel the tunneling current, $I_{\rm T}=0$. When $I^{\rm u}=I^{\rm l} $, mimicking the higher-dimensional case, we find a transverse resistance $R_{xy}=V_{\rm H}/I$, where $I=I^{\rm u}+I^{\rm l} $, of
\beq
R_{xy}= \frac{\pi q_{\rm B}}{e^2(2k_{\rm F}^{u})^3} \frac{\prod_{\sigma=\pm 1}{\left[ \left( \Delta k_{\rm F}  -\sigma q_{\rm B}\right)^2+l_{\rm br}^{-2}\right]} }{\Delta k_{\rm F} ^2 + q^2_{\rm B}+l_{\rm br}^{-2}}
\eeq
at $|\Delta k_{\rm F}|,q_{\rm B},l^{-1}_{\rm br} \ll k_{\rm F}^{\rm u}$ ($\Delta k_{\rm F}=k_{\rm F}^{\rm u} -k_{\rm F}^{\rm l}$). We make two observations: i) $R_{xy}$ is nonlinear in $B$ on the scale $\Delta B\sim {\rm max} \{|\Delta k_{\rm F} |/ed, (edl_{\rm br})^{-1}\}$, as illustrated in Fig.\ \ref{fig2}; ii) the `differential Hall coefficient' $ dR_{xy}/dB|_{B=0} =R_{\rm H}^{(0)}\times [\Delta k_{\rm F}^2+1/l_{\rm br}^2]/(2 k_{\rm F}^{\rm u})^2 $ is  suppressed below the Hall coefficient  $R^{(0)}_{\rm H}=-1/en_{\rm 2D}$  that one would expect in a two-dimensional electron gas.  Here, $n_{\rm 2D}=(n^{\rm u}+n^{\rm l})/d$ is an effective two-dimensional electron density between the two wires with one-dimensional densities $n^{\mu}=2 k_{\rm F}^{\mu}/\pi$.  Also the Hall response $R^{(-)}_{xy}$  to a difference $I^{(-)} =I^{\rm u}- I^{\rm l}$ between the currents through the wires,    $R^{(-)}_{xy}=-\pi q_{\rm B} \Delta k_{\rm F}/e^2 (\Delta k_{\rm F}^2+q_{\rm B}^2+l_{\rm br}^{-2}) $ (again at $|\Delta k_{\rm F}|,q_{\rm B},l^{-1}_{\rm br} \ll k_{\rm F}^{\rm u}$), where $V_{\rm H}=R_{xy} I+R^{(-)}_{xy} I^{(-)}$, is nonlinear in $B$ on the scale $\Delta B$. The differential Hall response to a difference in currents $dR^{(-)}_{xy}/dB|_{B=0} =[-8 \Delta k_{\rm F} k_{\rm F}^3/ (\Delta k_{\rm F}^2 +l_{\rm br}^{-2})^2]\times d R_{xy}/dB|_{B=0} $, however, is strongly enhanced.  Other mechanisms for the lifting of momentum conservation  are described  by  Eq.\ (\ref{LL})   with a (possibly) different $f$.  Both of our main conclusions hold for any kind of translational invariance breaking and also in the regime $v_{\rm s}/l_{\rm br}\gg eV_{\rm T}, I^{\rm u}/e,  I^{\rm l}/e \gg kT$. 

 {\em One spin-incoherent wire:} We next discuss the situation that  the upper wire  has a low electron density, $ k_{\rm F}^{\rm u} <  k_{\rm F}^{\rm l}$, and exhibits spin-incoherent physics, $kT\gg J^{\rm u}$, while the lower wire is still described by a conventional Luttinger liquid, $kT\ll J^{\rm l}$. This is motivated by the experiment of  Ref.\ \cite{auslaender:sci05}, where the observed loss of momentum conservation was attributed to only  one of the two wires.    We model the spin-incoherent  upper wire following Refs.\ \cite{fiete:prl04,kindermann:prl06b}. 
  Its  Green function  after   the spin trace  takes the form   \cite{fiete:prl04,kindermann:prl06b} 
   \bea \label{G}
 G_{{\rm u}\sigma}^>(x,x',\tau) &=& -i\int  {\frac{d\xi}{2\pi}dk\,  p_\sigma^{|k|}e^{i\xi k}}    \\
 &&\times \langle e^{-i \xi N_{x}(\tau)} c^\dag(x,\tau) c(x',0)e^{i\xi N_{x' }(0)}\rangle, \nonumber
\eea
and similarly for $G^<_{{\rm u}\sigma}$. Here, $c$ are spinless fermions that form a Luttinger liquid with  interaction parameter $g^{\rm u} <1$ inside the wire and $N_x$ is the number of fermions $c$ to the right of point $x$. We describe a current-carrying spin-incoherent wire of finite length $L$  contacted by noninteracting leads  following Ref.\  \cite{matveev:prl04} and evaluate Eq.\ (\ref{G}) by bosonization of the fermions $c$. Via the $x$-dependence of $N_x$ the  integrations in Eq.\ (\ref{G}) generate a space dependence of fermionic amplitudes on the scale $ (k_{\rm F}^{\rm u})^{-1}$. Since with our bosonization approach we access only the long wavelength limit we assume   that a magnetic field is applied in the plane of the wires that favors one of the  spin states, $1-p_\uparrow \ll 1$. The space dependence   in Eq.\ (\ref{G}) is then on the length scale  $ (k^{\rm u}_{\rm F}\ln p_\uparrow)^{-1}\gg(k_{\rm F}^{\rm u})^{-1}$. We expect, however, all results to remain qualitatively valid also at  $p_\uparrow\approx p_\downarrow $. We only evaluate  $G_{{\rm u }\uparrow}$ here since the minority spin tunnel current is expected to be negligible.  

In the following we address the regime of moderately low voltages, $kT, v_{\rm F}^{\rm u}/L \ll eV \ll \ln p_\uparrow /\delta$ with $\delta\sim 1/v_{\rm F}^{\rm u}k_{\rm F}^{\rm u}$. In this regime we obtain
\bea \label{Gincoh}
 &&G_{{\rm u}\uparrow}^>(x,x',\tau) = \frac{n^{\rm u} e^{i \pi I^{\rm u}(x-x')/e v_{\rm F}^{\rm u}}}{\sqrt{2\pi g\ln [(i \tau+\delta)/\delta]}}  \, \left(\frac{\delta}{i \tau+\delta}\right)^{1/2g^{\rm u}}\\
 &&\!\!\!\!\!\!\!\mbox{}\times \int dk\, p_\uparrow^{|k|} \cos \pi k \,e^{ -\pi^2[k - I^{\rm u}\tau/e-  (x-x')n^{\rm u} ]^2/2g\ln [(i \tau+\delta)/\delta]}\nonumber
 \eea
 [at $1/kT \gg  \tau\sim 1/eV \gg  \delta/\ln p_\uparrow$], where  now $n^{\rm u}=k_{\rm F}^{\rm u}/\pi$. As a consequence of spin-incoherence,  $G_{\rm u}$ decays quickly as a function of $x-x'$. Assuming that this is the dominant  mechanism for the lifting of   momentum conservation, ${\rm max}\{1/k^{\rm u}_{\rm F}\ln p_\uparrow,\sqrt{-g\ln eV_{\rm T}\delta}/k^{\rm u}_{\rm F}\} \ll l_{\rm br}$,  we then find from Eqs.\  (\ref{Gincoh}) and  (\ref{tunnel}) that 

\begin{widetext}
\beq \label{onetunnel}
I_{\rm T} \sim \sum_{\sigma^{\rm u},\sigma^{\rm l}=\pm 1}\frac{\ln p_\uparrow}{\ln^2 p_\uparrow + \pi^2[\sigma^{\rm u}+(q_{\rm B}/k_{\rm F}^{\rm u})+(\sigma^{\rm l} k_{\rm F}^{\rm l}/k_{\rm F}^{\rm u})]^2 } \left[ -V_{\rm T} + \frac{\pi I^{\rm u}}{e^2} \left(\frac{q_{\rm B}}{k_{\rm F}^{\rm u}}+\frac{\sigma^{\rm l} k_{\rm F}^{\rm l}}{k_{\rm F}^{\rm u}}\right)-\sigma^{\rm l} \frac{\pi I^{\rm l}}{2e^2}\right]^{\alpha_{\sigma^{\rm l}}}
\eeq
\end{widetext}
with the scaling exponents
\beq
\alpha_\sigma= \frac{1}{2g^{\rm u}} +\frac{g^{\rm u}}{2}\left(\frac{q_{\rm B}}{k_{\rm F}^{\rm u}}+\frac{\sigma k_{\rm F}^{\rm l}}{k_{\rm F}^{\rm u}}\right)^2-1 + \sum_{\nu\in\{{\rm c,s}\}}\frac{1}{4g^{\rm l}_{\nu}} +\frac{g^{\rm l}_{\nu}}{4}.
\eeq
In our limit $1-p_\uparrow\ll 1$, the first factor in Eq.\ (\ref{onetunnel}) consistently suppresses  large momentum transfers  $q_{\rm B}+\sigma^{\rm u} k_{\rm F}^{\rm u}+\sigma^{\rm l} k_{\rm F}^{\rm l}$, where our bosonization calculation is unreliable. For simplicity we now assume that the denominator $\ln^2 p_\uparrow + \pi^2[\bar{\sigma}^{\rm u}+(q_{\rm B}/k_{\rm F}^{\rm u})+(\bar{\sigma}^{\rm l} k_{\rm F}^{\rm l}/k_{\rm F}^{\rm u})]^2 $ of   the summand in Eq.\ (\ref{onetunnel})  with $\bar{\sigma}^{\rm u},\bar{\sigma}^{\rm l} =\pm 1$  is much smaller than the denominators in all other summands such that all but this one summand may be neglected.  

 We first note that, in contrast with the conventional Luttinger liquid, the tunneling current as a function of the applied voltages obeys a power law with an exponent $\alpha_{\bar{\sigma}}$ that depends on the  magnetic field $B$.   The $B$-dependence of $\alpha_{\bar{\sigma}}$ is due to a Fermi-edge singularity \cite{mahan:pr67,nozieres:pr69} with scattering phase shift $\delta \varphi=(\bar{\sigma}^{\rm u} k_{\rm F}^{\rm u}+\bar{\sigma}^{\rm l} k_{\rm F}^{\rm l}+q_{\rm B})/n^{\rm u}$.  To understand the origin of this phase shift we analyze  the tunneling rate, given by  amplitudes for the addition of an electron to the wire multiplied by  complex conjugated amplitudes, describing the removal of an electron. As a consequence of spin-incoherence, these pairs of amplitudes  are constrained to add and remove a spin at the same site of the spin configuration of the Wigner crystal (otherwise the spin expectation values  are suppressed by powers of $p_\uparrow$). Suppose that an electron  in the Wigner crystal crosses the point of tunneling
  during the time between   the addition and the removal of a tunneling electron. This shifts the spin background  by one lattice site. The above constraint can thus only be satisfied if the locations for the addition and the removal of the tunneling electron in space differ by the inter-electron distance $\Delta x=1/n^{\rm u}$. The  phase  $(\bar{\sigma}^{\rm u} k_{\rm F}^{\rm u}+\bar{\sigma}^{\rm l} k_{\rm F}^{\rm l}+q_{\rm B}) \Delta x$ that the tunneling electron picks up as a result translates into the effective phase shift $\delta\varphi$ for the electron of the Wigner crystal that crossed the point of tunneling. 

When a current $I^{\rm u} $  flows through the upper wire (at $I^{l}=0$), the upper crystal slides at velocity $v^{\rm u}=I^{\rm u}/en^{\rm u}$. So does the point of tunneling, which makes  
the phase shift $\delta\varphi$  time-dependent and thus induces a (Hall) voltage between the  wires.  As before we find from  Eq.\  (\ref{onetunnel}) that
\beq \label{Hall1}
V_{\rm H}=\left( B R_{\rm H}+ R'_{xy}\right)I^{\rm u}.
\eeq 
The first term in Eq.\ (\ref{Hall1}) remarkably describes a conventional Hall effect as known from higher dimensions with  $R_{\rm H}=R^{(0)}_{\rm H}$ at $n_{\rm 2D}=n^{\rm u}/d$ ($n^{\rm l}$ does not enter $n_{\rm 2D}$ since  the lower wire does not participate in the Hall effect). The second contribution to $V_{\rm H}$, proportional to $R'_{xy}=-\bar{\sigma}^{\rm l} k_{\rm F}^{\rm l}/e^2n^{\rm u}$, resembles the  anomalous Hall resistance in ferromagnets and does not vanish at $B=0$. Its origin is best understood in the reference frame   comoving with the sliding Wigner crystal in the upper wire. In that frame the energies of the electrons at the two Fermi points $\sigma^{\rm l} = \pm 1$ of the lower wire are shifted relative to those in the rest frame by $ v^{\rm u} \sigma^{\rm l} k_{\rm F}^{\rm l}$  through a Galilean boost. The resulting shift in chemical potential results in the extra voltage described by $R'_{xy}$. Note that Eq.\ (\ref{Hall1}) is invalid in zero magnetic field since our above assumption that one summand in Eq.\ (\ref{onetunnel}) dominates cannot be satisfied. In zero magnetic field one finds $R'_{xy}=0$, so  no anomalous Hall effect as in ferromagnets can be observed in this system. Current flow in the lower wire does not modify the Hall coefficient, but only changes  $R'_{xy}$.  

 {\em Two spin-incoherent wires:} We now analyze the situation that both wires are spin-incoherent, $kT\gg J^\mu$. 
At low voltages $|\ln(eV\delta)|\gg \pi^2/g(\ln p_\uparrow)^2$, $kT\ll eV$, we have
\bea \label{twotunnel}
&&I_{\rm T} \sim \sum_{\sigma^{\rm u},\sigma^{\rm l}=\pm 1}\frac{\ln p_\uparrow}{\ln^2 p_\uparrow + \pi^2[\sigma^{\rm u}+\bar{g}q_{\rm B}/g^{\rm u}_{\phantom{1}} k_{\rm F}^{\rm l} ]^2 } \nonumber \\
&& \mbox{}\times \{ {\rm u} \leftrightarrow {\rm l} \} \times \left[ -V_{\rm T} +q_{\rm B} \bar{g}\left( \frac{\pi I^{\rm u}}{e^2g_{\phantom{1}}^{\rm u} k_{\rm F}^{\rm l}}+ \frac{\pi I^{\rm l}}{e^2 g^{\rm l}_{\phantom{1}}k_{\rm F}^{\rm u}}\right)\right]^{\alpha } \nonumber\\
\eea
with $\bar{g}=g^{\rm u}_{\phantom{1}} g^{\rm l}_{\phantom{1}}n^{\rm u}n^{\rm l} / [g^{\rm u}_{\phantom{1}} (n^{\rm l} )^2+g^{\rm l}_{\phantom{1}} (n^{\rm u})^2 ]$ and $\alpha= 1/2g^{\rm u} + 1/2g^{\rm l} +\bar{g} q^2_{\rm B}/2 k_{\rm F}^{\rm u}k_{\rm F}^{\rm l}-1$.
We     find   
\beq \label{Hall2}
V_{\rm H} = B\left[R_{\rm H} I_{\phantom{1}} +  R^{(-)}_{\rm H} I^{(-)}_{\phantom{1}}\right]
\eeq
\cite{footnote2}. Unlike Eq.\ (\ref{Hall1}), that was derived under a $B$-dependent condition that allowed to neglect terms in Eq.\ (\ref{onetunnel}), Eq.\ (\ref{Hall2}) predicts a $V_{\rm H}$ linear in $B$ in the entire range of validity of our bosonization approach (set by the scale ${\rm min}\{k_{\rm F}^{\rm u},k_{\rm F}^{\rm l}\}$). This contrasts clearly with the conventional Luttinger liquid regime, where $V_{\rm H}$ becomes nonlinear on the scale $\Delta k_{\rm F}$, as shown in Fig.\ \ref{fig2}.  The Hall coefficient $R_{\rm H} =-(\bar{g} d/2e)(1/g^{\rm u} n^{\rm l}+1/g^{\rm l} n^{\rm u})$     is again of the order of the classically expected one and thus strongly enhanced compared to the conventional Luttinger liquid  (see Fig.\ \ref{fig2}). The magnitude of the Hall response to the difference between the currents through the two wires $R^{(-)}_{\rm H} =-(\bar{g} d/2e)(1/g^{\rm u} n^{\rm l}-1/g^{\rm l} n^{\rm u})$ is now smaller than $R_{\rm H}$, while it had been found to be strongly enhanced in the absence of spin-incoherence.  Counter-intuitively, the Hall response to currents in the wire with the lower electron density (found as $R_{\rm H}\pm R^{(-)}_{\rm H}$ with the positive sign if the upper wire has smaller density than the lower wire) is smaller than the one in the wire with higher density - although the lower density crystal slides faster  and experiences a stronger Lorentz force at $I^{\rm u}=I^{\rm l}$. 
  The conventional relation $V_{\rm H}=R_{\rm H}^{(0)} I$ holds  only  if both crystals slide at the same velocity $v^{\rm u}=I^{\rm u}/en^{\rm u}=I^{\rm l}/en^{\rm l}=v^{\rm l}$.  Also these features are readily understood by analyzing  the rate of tunneling between the wires.   The addition and the removal of an electron in each pair of amplitudes that contributes to it typically occur within a time $t_{\rm T} \sim 1/eV_{\rm T}$. Spin-incoherence again constrains the two amplitudes for adding and removing a spin to act at the same site of the spin configuration of each wire.     If $v^{\rm u}\neq v^{\rm l}$, however,   the spin configurations of the two wires  are   diverging in space at the average speed $v^{\rm u}-v^{\rm l}$. After the time $t_{\rm T}$ they can be aligned only if   the two crystals are compressed by amounts $\Delta x^{\rm u} $ and  $\Delta x^{\rm l} $ with $\Delta x^{\rm u}-\Delta x^{\rm l}= -(v^{\rm u}-v^{\rm l}) t_{\rm T} $. This costs an elastic energy $\epsilon_{\rm elastic}\propto (n^{\rm u}\Delta x^{\rm u})^2/g^{\rm u}+(n^{\rm l}\Delta x^{\rm l})^2/g^{\rm l}$.  
 Maximizing the probability $\exp(-S)$ of the corresponding deformation, where $S\propto \epsilon_{\rm elastic}$, under the constraint   $\Delta x^{\rm u}-\Delta x^{\rm l}= -(v^{\rm u}-v^{\rm l}) t_{\rm T} $ we find $\Delta x^{\rm u}=-t_{\rm T}(v^{\rm u}-v^{\rm l})\bar{g} n^{\rm l}/n^{\rm u}g^{\rm l}$. This distortion of the crystals results in a modified effective velocity of an electron during the tunneling process of $v_{\rm eff}=- (R_{\rm H} I_{\phantom{1}} +  R^{(-)}_{\rm H} I^{(-)}_{\phantom{1}})/d$. The corresponding Lorentz force implies Eq.\ (\ref{Hall2}).   Now the reason for the suppression of the Hall coefficient of the low-density wire noted above is evident: because the electron configuration in the low-density wire is  deformed more easily  $v_{\rm eff}$ (and thus $V_{\rm H}$) is predominantly determined by the wire with the higher density and depends only weakly on  the current through the low-density wire.
 
  {\em Conclusions:} We have studied  tunneling   between parallel quantum wires at low electron density. An almost conventional Hall effect has been shown to emerge as the wires enter the spin-incoherent regime of small spin bandwidth. The Hall coefficient is of the order of the one classically expected at a given electron density and the Hall voltage only weakly depends on the difference of the currents through the two wires. In contrast, two wires in the absence of spin-incoherence with weak translational symmetry breaking, $\Delta k_{\rm F} l_{\rm br} \gtrsim 1$, have a  Hall coefficient that is suppressed under its classical value by a factor of $(\Delta k_{\rm F}/k_{\rm F})^2$, where   $\Delta k_{\rm F}$ is the difference between the Fermi wavevectors of the two wires with average wavevector $k_{\rm F}$, while the  Hall response to a difference between the currents that flow through such wires  is anomalously enhanced by a factor $( k_{\rm F}/\Delta k_{\rm F})^3$ compared to the response to the average current. Moreover, wires in the conventional regime exhibit a nonlinear magnetic field dependence  on the scale set by  $\Delta k_{\rm F}$ (again for $\Delta k_{\rm F}l_{\rm br} \gtrsim 1$). In contrast, spin-incoherent conductors are predicted to produce a   transverse voltage that is linear in the magnetic field up to a scale of the order of the Fermi wavevectors themselves. 
  This together with magnetic field dependent tunneling exponents clearly identifies  spin-incoherent physics in experiments like those of Refs.\ \cite{auslaender:sci02,auslaender:sci05}. In particular, it distinguishes spin-incoherence  from the  effects of disorder. Such measurements are thus a very promising avenue in the search for this novel  regime of interacting quantum wires.
  
  The author thanks very much P. W. Brouwer and A. Yacoby for discussions of the results and valuable remarks.
 

\vspace{-.5cm}
 \begin{thebibliography}{22}
  \vspace{-.5cm}

\expandafter\ifx\csname natexlab\endcsname\relax\def\natexlab#1{#1}\fi
\expandafter\ifx\csname bibnamefont\endcsname\relax
  \def\bibnamefont#1{#1}\fi
\expandafter\ifx\csname bibfnamefont\endcsname\relax
  \def\bibfnamefont#1{#1}\fi
\expandafter\ifx\csname citenamefont\endcsname\relax
  \def\citenamefont#1{#1}\fi
\expandafter\ifx\csname url\endcsname\relax
  \def\url#1{\texttt{#1}}\fi
\expandafter\ifx\csname urlprefix\endcsname\relax\def\urlprefix{URL }\fi
\providecommand{\bibinfo}[2]{#2}
\providecommand{\eprint}[2][]{\url{#2}}

\bibitem[{\citenamefont{Klitzing et~al.}(1980)\citenamefont{Klitzing, Dorda,
  and Pepper}}]{klitzing:prl80}
\bibinfo{author}{\bibfnamefont{K.~v.} \bibnamefont{Klitzing}},
  \bibinfo{author}{\bibfnamefont{G.}~\bibnamefont{Dorda}}, \bibnamefont{and}
  \bibinfo{author}{\bibfnamefont{M.}~\bibnamefont{Pepper}},
  \bibinfo{journal}{Phys. Rev. Lett.} \textbf{\bibinfo{volume}{45}},
  \bibinfo{pages}{494} (\bibinfo{year}{1980}).

\bibitem[{\citenamefont{Tsui et~al.}(1982)\citenamefont{Tsui, Stormer, and
  Gossard}}]{tsui:prl82}
\bibinfo{author}{\bibfnamefont{D.~C.} \bibnamefont{Tsui}},
  \bibinfo{author}{\bibfnamefont{H.~L.} \bibnamefont{Stormer}},
  \bibnamefont{and} \bibinfo{author}{\bibfnamefont{A.~C.}
  \bibnamefont{Gossard}}, \bibinfo{journal}{Phys. Rev. Lett.}
  \textbf{\bibinfo{volume}{48}}, \bibinfo{pages}{1559} (\bibinfo{year}{1982}).

\bibitem[{\citenamefont{Pugh and Rostoker}(1953)}]{pugh:rmp53}
\bibinfo{author}{\bibfnamefont{E.~M.} \bibnamefont{Pugh}} \bibnamefont{and}
  \bibinfo{author}{\bibfnamefont{N.}~\bibnamefont{Rostoker}},
  \bibinfo{journal}{Rev. Mod. Phys.} \textbf{\bibinfo{volume}{25}},
  \bibinfo{pages}{151} (\bibinfo{year}{1953}).

\bibitem[{\citenamefont{Sundaram and Niu}(1999)}]{niu:prb99}
\bibinfo{author}{\bibfnamefont{G.}~\bibnamefont{Sundaram}} \bibnamefont{and}
  \bibinfo{author}{\bibfnamefont{Q.}~\bibnamefont{Niu}},
  \bibinfo{journal}{Phys. Rev. B} \textbf{\bibinfo{volume}{59}},
  \bibinfo{pages}{14915} (\bibinfo{year}{1999}).

\bibitem[{\citenamefont{Fiete}(2006)}]{fiete:rmp07}
\bibinfo{author}{\bibfnamefont{G.~A.} \bibnamefont{Fiete}},
   \bibinfo{journal}{Rev. Mod. Phys. }
  \textbf{\bibinfo{volume}{79}}, \bibinfo{eid}{801} (\bibinfo{year}{2007}).

\bibitem[{\citenamefont{Kindermann}(2007)}]{kindermann07}
\bibinfo{author}{\bibfnamefont{M.}~\bibnamefont{Kindermann}},
   \bibinfo{journal}{Phys. Rev. Lett.}
  \textbf{\bibinfo{volume}{99}}, \bibinfo{eid}{076801} (\bibinfo{year}{2007}).
  
  

\bibitem[{\citenamefont{Tilahun and Fiete}(2007)}]{fiete07}
\bibinfo{author}{\bibfnamefont{D.}~\bibnamefont{Tilahun}} \bibnamefont{and}
  \bibinfo{author}{\bibfnamefont{G.~A.} \bibnamefont{Fiete}},
  \bibinfo{journal}{cond-mat/0706.3418}  (\bibinfo{year}{2007}).

\bibitem[{\citenamefont{Auslaender et~al.}(2002)\citenamefont{Auslaender,
  Yacoby, de~Picciotto, Baldwin, Pfeiffer, and West}}]{auslaender:sci02}
\bibinfo{author}{\bibfnamefont{O.~M.} \bibnamefont{Auslaender}},
  \bibinfo{author}{\bibfnamefont{A.}~\bibnamefont{Yacoby}},
  \bibinfo{author}{\bibfnamefont{R.}~\bibnamefont{de~Picciotto}},
  \bibinfo{author}{\bibfnamefont{K.~W.} \bibnamefont{Baldwin}},
  \bibinfo{author}{\bibfnamefont{L.~N.} \bibnamefont{Pfeiffer}},
  \bibnamefont{and} \bibinfo{author}{\bibfnamefont{K.~W.} \bibnamefont{West}},
  \bibinfo{journal}{Science} \textbf{\bibinfo{volume}{295}},
  \bibinfo{pages}{825} (\bibinfo{year}{2002}).

\bibitem[{\citenamefont{Auslaender et~al.}(2005)\citenamefont{Auslaender,
  Steinberg, Yacoby, Tserkovnyak, Halperin, Baldwin, Pfeiffer, and
  West}}]{auslaender:sci05}
\bibinfo{author}{\bibfnamefont{O.~M.} \bibnamefont{Auslaender}},
  \bibinfo{author}{\bibfnamefont{H.}~\bibnamefont{Steinberg}},
  \bibinfo{author}{\bibfnamefont{A.}~\bibnamefont{Yacoby}},
  \bibinfo{author}{\bibfnamefont{Y.}~\bibnamefont{Tserkovnyak}},
  \bibinfo{author}{\bibfnamefont{B.~I.} \bibnamefont{Halperin}},
  \bibinfo{author}{\bibfnamefont{K.~W.} \bibnamefont{Baldwin}},
  \bibinfo{author}{\bibfnamefont{L.~N.} \bibnamefont{Pfeiffer}},
  \bibnamefont{and} \bibinfo{author}{\bibfnamefont{K.~W.} \bibnamefont{West}},
  \bibinfo{journal}{Science} \textbf{\bibinfo{volume}{308}},
  \bibinfo{pages}{88} (\bibinfo{year}{2005}).

\bibitem[{\citenamefont{Fiete et~al.}(2005)\citenamefont{Fiete, Qian,
  Tserkovnyak, and Halperin}}]{fiete:prb05b}
\bibinfo{author}{\bibfnamefont{G.~A.} \bibnamefont{Fiete}},
  \bibinfo{author}{\bibfnamefont{J.}~\bibnamefont{Qian}},
  \bibinfo{author}{\bibfnamefont{Y.}~\bibnamefont{Tserkovnyak}},
  \bibnamefont{and} \bibinfo{author}{\bibfnamefont{B.~I.}
  \bibnamefont{Halperin}}, \bibinfo{journal}{Phys. Rev. B}
  \textbf{\bibinfo{volume}{72}}, \bibinfo{eid}{045315} (\bibinfo{year}{2005}).

\bibitem[{foo({\natexlab{a}})}]{footnote00}
\bibinfo{note}{A spin-polarizing magnetic field that can reverse the loss of momentum resolution if caused by spin-incoherence would be an another possibility. Since the orbital effects of that magnetic field are also able to reduce disorder effects, however, additional signatures are desirable.}








\bibitem[{\citenamefont{Tserkovnyak et~al.}(2003)\citenamefont{Tserkovnyak,
  Halperin, Auslaender, and Yacoby}}]{tserkovnyak:prb03}
\bibinfo{author}{\bibfnamefont{Y.}~\bibnamefont{Tserkovnyak}},
  \bibinfo{author}{\bibfnamefont{B.~I.} \bibnamefont{Halperin}},
  \bibinfo{author}{\bibfnamefont{O.~M.} \bibnamefont{Auslaender}},
  \bibnamefont{and} \bibinfo{author}{\bibfnamefont{A.}~\bibnamefont{Yacoby}},
  \bibinfo{journal}{Phys. Rev. B} \textbf{\bibinfo{volume}{68}},
  \bibinfo{pages}{125312} (\bibinfo{year}{2003}).

\bibitem[{foo({\natexlab{a}})}]{footnote0}
\bibinfo{note}{At small $\lambda$, when electrons are tightly confined to the
  wires, effects of the magnetic field on the orbital wave-functions
  \cite{kirczenow:prb88} and thus on $\lambda$ can be neglected.}

\bibitem[{\citenamefont{Haldane}(1981)}]{haldane:jpc81}
\bibinfo{author}{\bibfnamefont{F.~D.~M.} \bibnamefont{Haldane}},
  \bibinfo{journal}{J. Phys. C} \textbf{\bibinfo{volume}{14}},
  \bibinfo{pages}{2585} (\bibinfo{year}{1981}).

\bibitem[{\citenamefont{Giamarchi}(2004)}]{giamarchi:bo04}
\bibinfo{editor}{\bibfnamefont{T.}~\bibnamefont{Giamarchi}}, ed.,
  \emph{\bibinfo{title}{Quantum Physics in One Dimension}}
  (\bibinfo{publisher}{Oxford University Press}, \bibinfo{year}{2004}).

\bibitem[{\citenamefont{Fiete and Balents}(2004)}]{fiete:prl04}
\bibinfo{author}{\bibfnamefont{G.~A.} \bibnamefont{Fiete}} \bibnamefont{and}
  \bibinfo{author}{\bibfnamefont{L.}~\bibnamefont{Balents}},
  \bibinfo{journal}{Phys. Rev. Lett.} \textbf{\bibinfo{volume}{93}},
  \bibinfo{eid}{226401} (\bibinfo{year}{2004}).

\bibitem[{\citenamefont{Kindermann et~al.}(2006)\citenamefont{Kindermann,
  Brouwer, and Millis}}]{kindermann:prl06b}
\bibinfo{author}{\bibfnamefont{M.}~\bibnamefont{Kindermann}},
  \bibinfo{author}{\bibfnamefont{P.~W.} \bibnamefont{Brouwer}},
  \bibnamefont{and} \bibinfo{author}{\bibfnamefont{A.~J.}
  \bibnamefont{Millis}}, \bibinfo{journal}{Phys. Rev. Lett.}
  \textbf{\bibinfo{volume}{97}}, \bibinfo{pages}{036809}
  (\bibinfo{year}{2006}).

\bibitem[{\citenamefont{Matveev}(2004)}]{matveev:prl04}
\bibinfo{author}{\bibfnamefont{K.~A.} \bibnamefont{Matveev}},
  \bibinfo{journal}{Phys. Rev. Lett.} \textbf{\bibinfo{volume}{92}},
  \bibinfo{eid}{106801} (\bibinfo{year}{2004}).

 

\bibitem[{\citenamefont{Mahan}(1967)}]{mahan:pr67}
\bibinfo{author}{\bibfnamefont{G.~D.} \bibnamefont{Mahan}},
  \bibinfo{journal}{Phys. Rev.} \textbf{\bibinfo{volume}{163}},
  \bibinfo{pages}{612} (\bibinfo{year}{1967}).

\bibitem[{\citenamefont{Nozi\`eres and De~Dominics}(1969)}]{nozieres:pr69}
\bibinfo{author}{\bibfnamefont{P.}~\bibnamefont{Nozi\`eres}} \bibnamefont{and}
  \bibinfo{author}{\bibfnamefont{C.~T.} \bibnamefont{De~Dominics}},
  \bibinfo{journal}{Phys. Rev.} \textbf{\bibinfo{volume}{178}},
  \bibinfo{pages}{1097} (\bibinfo{year}{1969}).

\bibitem[{foo({\natexlab{c}})}]{footnote2}
\bibinfo{note}{Eq.\ (\ref{Hall2}) does not require the condition
  $-\ln(eV\delta)\gg \pi^2/g(\ln p_\uparrow)^2$, but only assumes
  $-\ln(eV\delta)\gg 1$}.

\bibitem[{\citenamefont{Kirczenow}(1988)}]{kirczenow:prb88}
\bibinfo{author}{\bibfnamefont{G.}~\bibnamefont{Kirczenow}},
  \bibinfo{journal}{Phys. Rev. B} \textbf{\bibinfo{volume}{38}},
  \bibinfo{pages}{10958} (\bibinfo{year}{1988}).

\end{thebibliography}
 \end{document}